# Suggestion on Quantum Computer Circuits

Helio V. Fagundes

*Instituto de Física Teórica, Universidade Estadual Paulista*

Rua Dr. Bento Teobaldo Ferraz, 271, bloco 2, São Paulo, SP 01140-070, Brazil

E-mail address: helio@ift.unesp.br

ABSTRACT

For quantum computer circuits, it is proposed that they have, besides the presently used compact graphs, an expanded system of subgraphs, in line with the quantum mechanics superposition axiom. The representation of each process by these suggested graphs is equivalent to the algebraic notation of the process

1.  INTRODUCTION

As Dale and Lewis (2011) emphasize in their textbook on computer science, there are three equivalent representations for classical computer processes: the Boolean, the graphical, and the truth table ones. As for quantum computations, we have an algebraic complete notation, and a graphical one which, because of quantum superposition and entanglement, is not equivalent to the algebraic one.

Here I want to suggest a new class of graphical representation for quantum computer circuits, which is equivalent to their algebraic representation. It is composed of multiple subgraphs for a given process, mimicking the quantum mechanical superposition axiom. Each of these subgraphs represents a non-entangled tensor product of qubits, with each line corresponding to a single, definite qubit – except for output lines, which sometimes it is more convenient to be kept in entangled form, as in the second example below.

My starting point is to make the graphs show the quantum mechanical superposition principle;

due to the number of possibilities, they are often labeled with the corresponding qubits in ket notation. The idea is not to totally replace the presently used graphs – which I will call *compact* – but also to have a more explicit representation of the processes. The gates or operators in the compact graph are replaced by their effective action in each subgraph. The geometric structure of each subgraph is essentially the same as that of the corresponding compact graph, with the operations and state labels making the difference. As a bonus, the subgraphs provide a graphically appealing way for reading out the final result, and for checking over the intermediate steps for errors. This is hoped to help the learning of novices to quantum computation, who often are also novices to quantum mechanics. In each of the next three sections I illustrate this idea with an example from quantum computer theory.

## 2. THE CNOT GATE

As a first example, Fig. 1 shows the compact graph for the CNOT gate − see, for example, Portugal et al (2012), Nakahara and Ohmi (2011), Chuang and Nielsen (2000), for quantum computer theory − on the left-hand side, and the two subgraphs for the same gate on the right-hand side, which constitute what will here be called *extended* graphs. The *addition mod* 2 operator is replaced by $I$ in the first subgraph, and by $X$ in the second one, as these are its effective actions in the subgraphs. Thus the sum of subgraphs in Fig.1 allows the computation of the output, being equivalent to the algebraic computation – which is not the case for the compact, entangled graph (cf. Portugal et al 2012, Sec. 2.1).

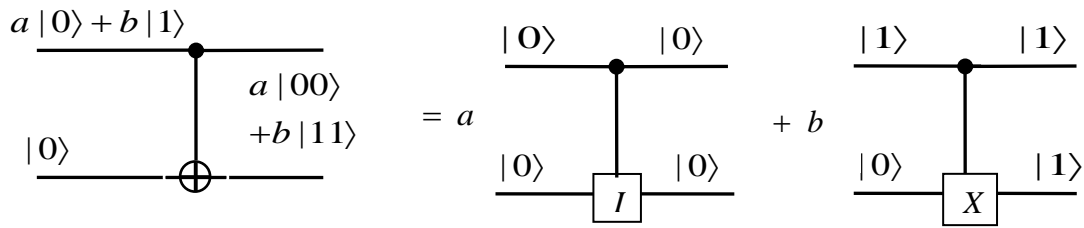

**Fig. 1** The compact graph for CNOT gate is equivalent to a sum of
two subgraphs in its extended representation. $\oplus$ is the addition mod 2
operation, $I$ is the identity, and $X$ is Pauli $\sigma_x$ matrix

## 3. SHOR'S QECC

The second example refers to Shor's nine-qubit quantum error correction code (QECC) – cf. Shor (1995). Here I am mostly following the development in Nakahara and Ohmi (2011), Sec. 10.3. I rewrite the QECC encoding circuit using what might be called a *subroutine* circuit (SRC), developed in Fig. 2 as an extended graph. The subroutine obtained in Fig. 2 is represented as the 'black box' circuit SRC in Fig. 3, where $|\psi_\pm\rangle \equiv (|000\rangle \pm |111\rangle)/\sqrt{2}$. These results are then applied to Fig. 4, which shows the encoding circuit for this QECC, making use of the SRC subroutine. Note that this subroutine strategy may also be done with higher level subprograms, to avoid a 'huge number of graphs,' as an anonymous referee put it.

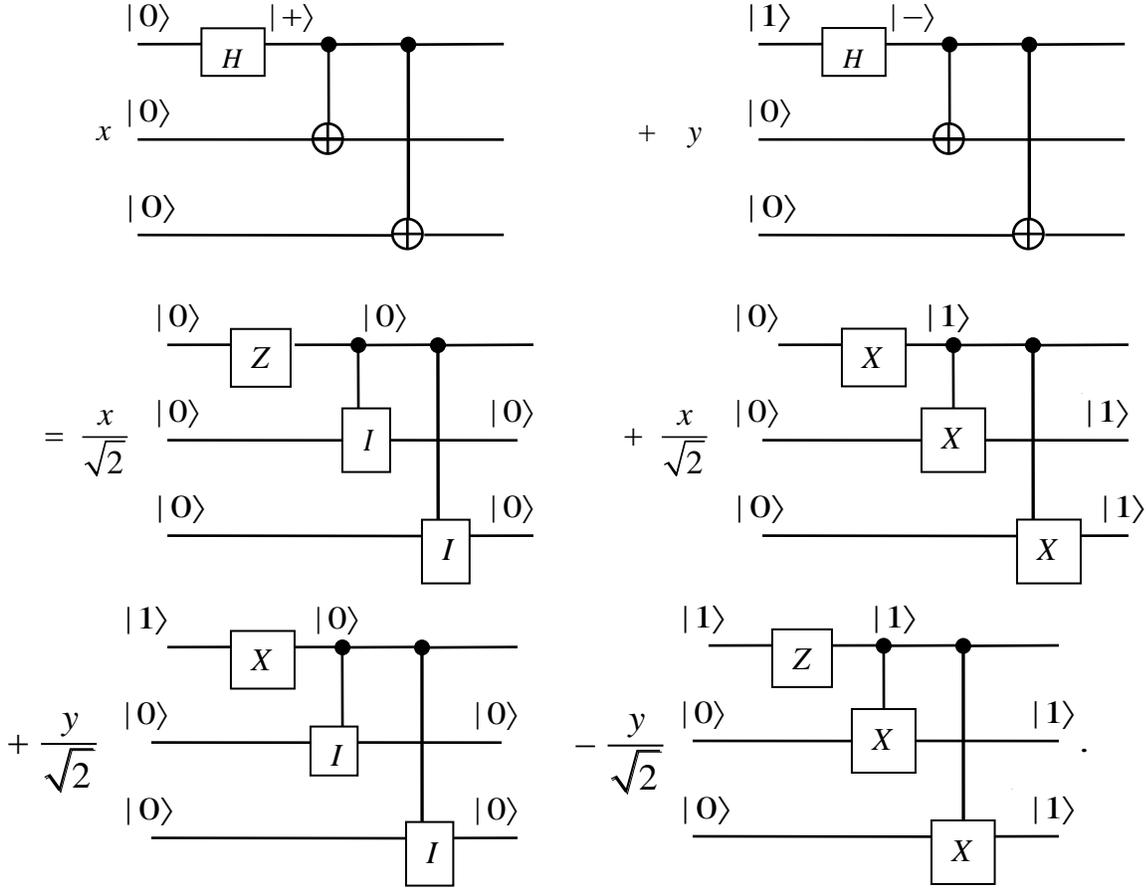

**Fig. 2** A subroutine to be included in Shor's 9-qubit code. $H = (X+Z)/\sqrt{2}$ is Hadamard gate, with Pauli matrices $X = \sigma_x$, $Z = \sigma_z$; $|\pm\rangle \equiv (|0\rangle \pm |1\rangle)/\sqrt{2}$. The output is $x|\psi_+\rangle + y|\psi_-\rangle$, where $|\psi_\pm\rangle \equiv (|000\rangle \pm |111\rangle)/\sqrt{2}$

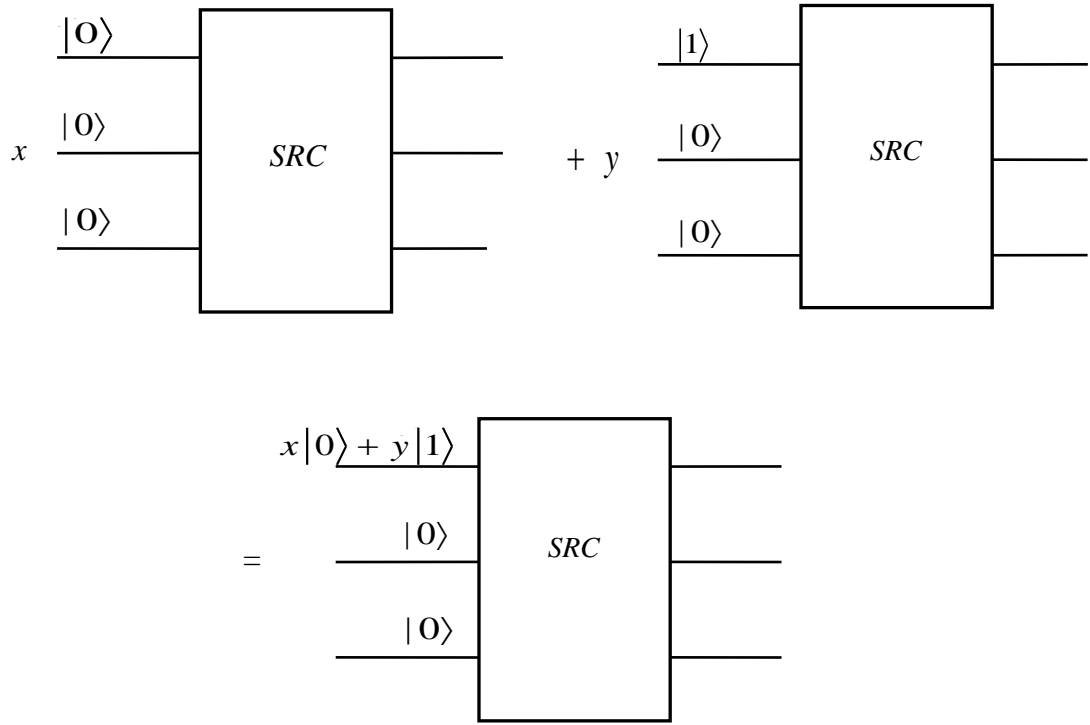

**Fig. 3** The preparation subroutine circuit for Shor's QECC in terms of the SRC 'black box,' which is the preparation circuit in Fig. 2. With $|\psi_\pm\rangle \equiv (|000\rangle \pm |111\rangle)/\sqrt{2},$ the outputs are $x|\psi_+\rangle$ and $y|\psi_-\rangle$ for the top graphs, and their sum for the bottom one

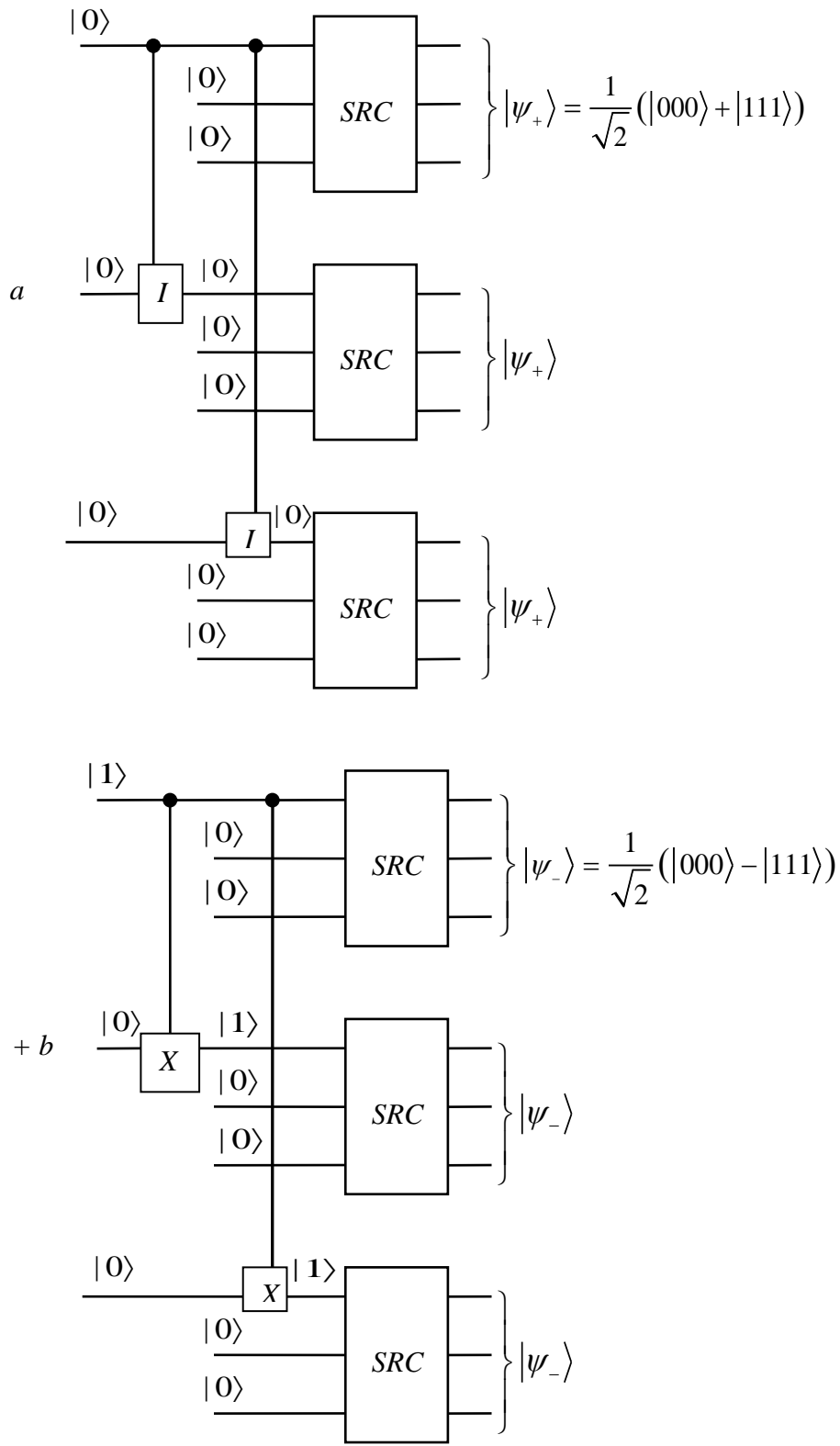

**Fig. 4** Encoding circuit for Shor's QECC. The output is $a|\psi_+\rangle^{\otimes 3} + b|\psi_-\rangle^{\otimes 3}$

## 4. AMPLITUDE DAMPING

The third and last example is derived from Exercise 8.20 in Nielsen and Chuang (2000). Here their top line initial density operator $\rho_{in}$ is replaced in Fig. 5 by initial qubit $|\psi_{in}\rangle$ and the original circuit in the exercise is adapted to produce an entangled tensor output for the complete system. Their final measurement in the lower line is omitted, for we are not discussing the tracing out of the environment. In this paper's context of subgraphs the total system processes are represented in Fig. 5. The initial state is $|\Psi_{in}\rangle = |\psi_{in}\rangle \otimes |0\rangle = \alpha|00\rangle + \beta|10\rangle$, and the authors reach the output $|\Psi_{out}\rangle = \alpha|00\rangle + \beta[\cos(\theta/2)|10\rangle + \sin(\theta/2)|01\rangle]$ by physical reasoning. The purpose of the exercise is to verify that their circuit, analogous to Fig. 5 here, produces this result, as may be confirmed by inspection of the figure.

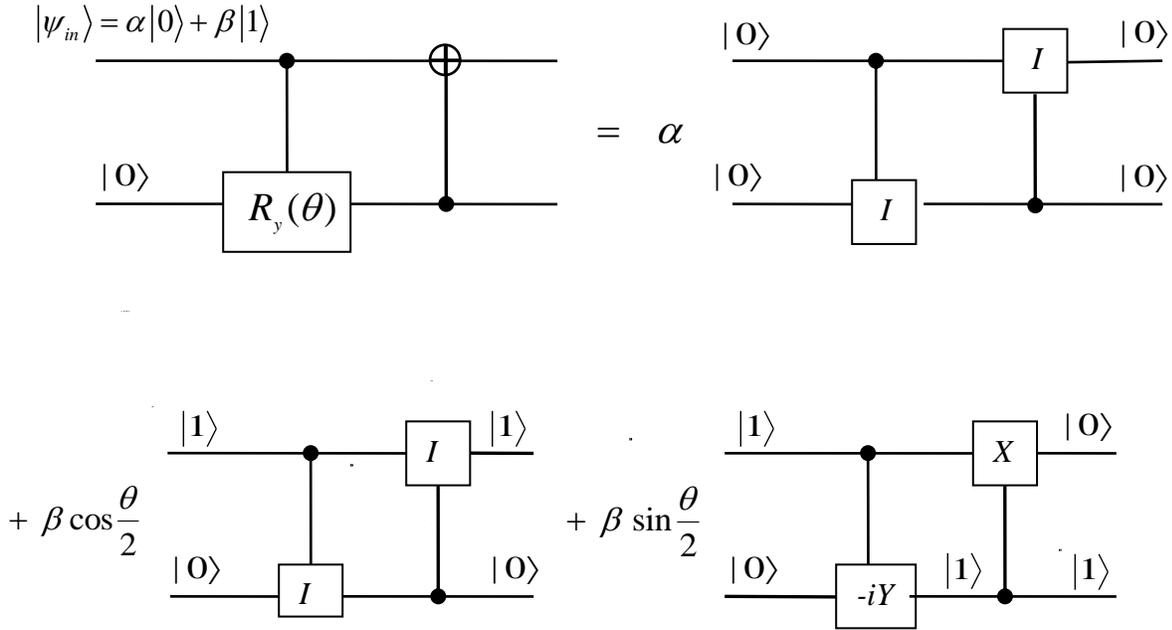

**Fig. 5** A circuit for producing an amplitude damping channel. The top line represents the principal system, the other one represents the environment. Symbols $X \equiv \sigma_x$, $Y \equiv \sigma_y$ are Pauli spin matrices and $R_y(\theta) = I\cos(\theta/2) - iY\sin(\theta/2)$ is a rotation by angle $\theta$ around the y-axis

The reader will have noted that the sum of subgraphs which constitute an extended graph for a process has the structure of the quantum mechanical superposition principle, and is equivalent to the linear-algebraic description of the same process.

______________________________________

.